\documentstyle[11pt,paspconf,epsf]{article}

\begin{document}

\title{ESTIMATING MASSES IN CLUSTERS OF GALAXIES}

\author{H\'ector Aceves and Jaime Perea}
\affil{Instituto de Astrof\'{\i}sica de Andaluc\'{\i}a \\
Apdo. Postal 3004, Granada 18080. SPAIN \quad aceves@iaa.es}

\begin{abstract}
The projected (PME) and virial mass estimator (VME) are revisited and tested using an $N$-body equilibrium system. It is found that the PME can overestimate the mass by $\approx 40$\% if a cluster of galaxies is sampled only about its {\it effective} radius, $R_{\rm e}$. The exact value of this error depends on the mass profile of the cluster. The VME can yield errors of $\approx 20$\% if we are also able to sample about the $R_{\rm e}$. Both estimators yield the correct total mass for an equilibrium system. Theoretically, if we know the total extent of the system, the VME provides an accurate mass at different radii, provided its gravitational potential term is correctly taken into consideration. The VME can provide acceptable masses for equilibrium systems with anisotropic velocity dispersions.
\end{abstract}


\section{Introduction}

Clusters of galaxies provide an important tool to study the large-scale 
structure of the universe (e.g. Peebles 1980), and several methods are used to determine their total mass and mass profile. Recently, some questions regarding the accuracy and consistency of the PME and VME have been raised by several authors. 

Carlberg, Yee \& Ellingson (1997, hereafter CYE) found that the VME overestimates mass by about $\approx 20$\% in clusters of galaxies, and attribute this discrepancy to the neglect of a surface pressure ($3PV$) term in the continuous form of the virial theorem. This $3PV$ term is starting to be used as a correction term when mass estimates from the VME are calculated (e.g. Girardi, et al. 1998). Perhaps more critical is the situation in $N$-body simulations where there is no ambiguity regarding positions and velocities of particles, and the obtained masses using either the PME or the VME do not agree with the true mass (e.g. Thomas \& Couchman 1992).

Here we revisit the PME and the VME and test their consistency using an $N$-body equilibrium system, and explain the discrepancies that arise in the relevant literature. A more comprehensive treatment of these matters, along with a determination of the mass profile of  Coma cluster can be found elsewhere (Aceves \& Perea 1998). We consider here, however, the performance of the VME for  a system in equilibrium but with a non-isotropic velocity distribution.

\section{Projected and Virial Mass Estimators}

        For an isotropic and spherically symmetric gravitational system the PME and the VME are, respectively (e.g. Heisler et al. 1985; Perea, et al. 1990):
\begin{equation}
M_{\rm P} =\frac{1}{G} \frac{1}{N} \sum_i v_{z_i}^2 R_i =  \frac{32}{\pi G} 
\langle v_z^2 R \rangle  \;, \quad\quad
M_{\rm V} = \frac{3 \pi N}{2 G} \frac{\sum_i v_{z_i}^2}{\sum_{i<j} 1/R_{ij}} \,,
\end{equation}
where $v_{z_i}$ is the observed line-of-sight velocity of a galaxy relative to the cluster mean, $R_i$ is its projected radius from the center of the distribution, and $R_{ij}$  denotes the projected inter-galaxy separation.

If we consider partial sampling of a system, a correction term for the PME has to be included (e.g. Haller \& Melia 1996, Aceves \& Perea 1998), namely:
\begin{equation}
\Delta\equiv \frac{32}{\pi G} \langle v^2_zR \rangle - M(r) = \frac{8\pi}{G M(r)}
\,r^4\rho(r) \sigma_r^2(r) \;,
\end{equation}
where  $M(r)$ is the true mass of the system, $\rho (r)$ is the mass density and $\sigma_r (r)$ the radial velocity dispersion.
For the VME, CYE have suggested a surface pressure term $3PV\equiv 4 \pi r^3 \rho \sigma_r^2$ as a correction to virial mass estimates to avoid overestimation (see also Girardi, et al. 1998).  We will now test these estimators, and the suggested correction terms, for an $N$-body equilibrium model.

\begin{figure}
\plotfiddle{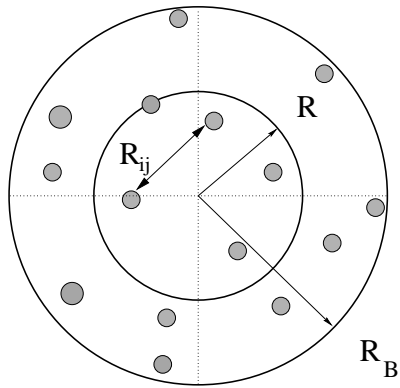}{5cm}{0}{33}{33}{-75}{0}
\caption{Schematic representation of the use of the PME and the VME by apertures in a cluster of galaxies; $R_{\rm B}$ is the physical boundary of the system, $R$ is the radius of the aperture, and $R_{ij}$ is the inter-particle separation between two galaxies.}
\end{figure}

        Throughout this work the units used are such that the total mass is $M=1$, the {\it effective} radius is $R_{\rm e}=1$, and $G=1$. A Monte-Carlo realization of $N=10^4$ particles for a Dehnen $\gamma=0$ model (Dehnen 1993) was used to test the mass estimators referred above. The effective radius in this model, i.e. the locus where the projected mass is one-half of the total mass, is $R_{\rm e}=2.9036a\,$, with $a$ being a particular scale radius. We use here Dehnen model as an illustrative case, due mainly to the simplicity of  its expression for the velocity dispersion, but we have considered other models and similar trends as those described here were found; although with different numerical values.

	In order to apply the PME and the VME we proceeded in a similar fashion as one would observationally do. We divided the complete system in different `apertures' of increasing radii, and applied both the PME and the VME just to the particles inside each of them; see Fig. 1. This approach mimics, particularly for the VME,  the lack of information one may have on the total extent of a real cluster of galaxies, especially if large amounts of dark matter exist outside their optical boundary.

        In Fig. 2, our results are shown. Figure 2a. shows the difference between $M_{\rm P}$ and $M(r)$, and with respect to $M(R)$; the theoretical value of $\Delta (r)$ is also displayed. In Fig. 2b. the differences of the VME applied by apertures, both in its projected and non-projected versions, with the true mass are shown. Almost perfect agreement with the theoretical expectations of Eq. (2) exists for the PME, except by the Poisson noise in the discrete $N$-body model. From the values of $\Delta (r)$ and the numerical experiments it follows that the PME may overestimates the true mass by $\approx 20$\% if sampling is made only around $\approx 1 R_{\rm e}$, and if the mass profile is similar to a Dehnen model. For profiles similar to Hernquist (1990) or De Vaucouleurs (1948), i.e. `cuspy', the error is smaller while for
systems of a larger core, e.g. a King Modified (see Aceves \& Perea 1998), it can be up to $\approx 40$\%.

\begin{figure}
\plotfiddle{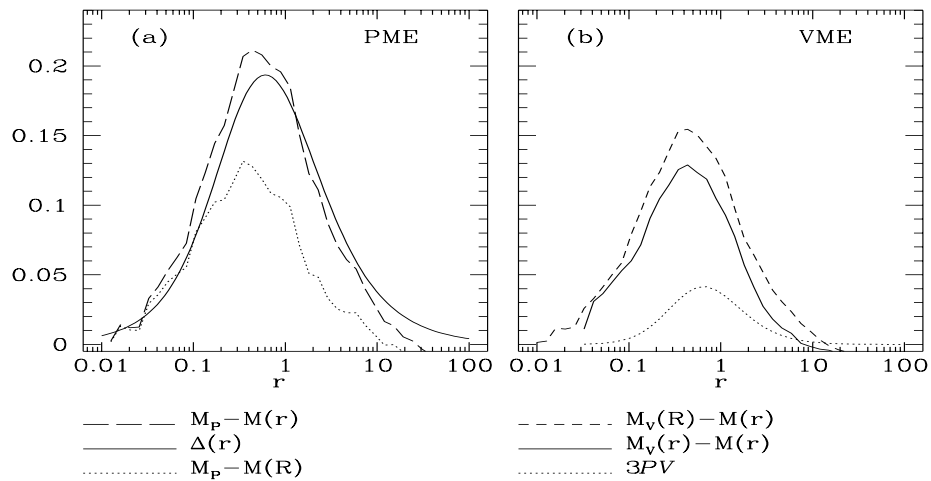}{6cm}{-90}{50}{35}{-210}{190}
\caption{Results of the application of the PME and VME to an $N$-body equilibrium system following a Dehnen ($\gamma=0$) profile.}
\end{figure}

        The discrepancy of the VME result applied by apertures with the true mass requires a somewhat different treat. CYE attribute this kind of error to a neglect of the $3PV$ term in the continuous virial theorem:
\begin{equation}
4 \pi r^3 \rho \sigma^2_r  - 12 \pi  \int^r r^2 \rho  \sigma_r^2 \,{\rm d}r = 
- 4 \pi G \int^r r \rho M(r) \,{\rm d}r   \;,
\end{equation}
where the {\it surface pressure} term is $3PV\equiv 4 \pi r^3 \rho \sigma_r^2\,$. We can readily estimate the position of the maximum of the $3PV$ term for a Dehnen model, where
\begin{equation}
\rho(r) = \frac{3 M a}{4 \pi}\, \frac{1}{(r+a)^4} \;\;,  \quad\quad
\sigma_r^2 (r)  = G M \, \frac{a + 6 r}{30 \,(r+a)^2} \;\;,
\end{equation}
obtaining  $\,r_{\rm max}=(7+\sqrt{65})a/8 \approx 0.65$, yielding a maximum value of $3PV_{\rm max} \approx 0.04\,$; see Fig. 2b. This value is about the maximum overestimate in mass when the $3PV$ term is not taken into consideration in (3).

        Moreover, without the $3PV$ term the resulting mass from Eq. 3 would be $M_{\rm NSP} (r) = 3 r \sigma_r^2(r)/G$, which overestimates the mass for $r \la R_{\rm e}$, the near isothermal region of the model, and underestimates it for $r \ga R_{\rm e}$. For example, we have that at $r=100 R_{\rm e}$ the mass is $M_{\rm NSP}\approx 0.6$ instead of $\,\approx 1\,$; i.e. about a 40\% underestimate error is obtained for the total mass.

The physical reason for the discrepancy in the aperture evaluation of the VME value with the true mass $M(r)$ is the incorrect evaluation of the potential energy term, i.e. that involving $\sum 1/R_{ij}$, on equation (1).
 	The previous summation has to be done for all $j$-th particles in the system, not only over those inside the aperture radius defined by the $i$-th particle (Aceves \& Perea 1998). When this is done, we find very good agreement between the calculated and theoretical masses at different radii; see Fig. 3.
This is basically a consequence of the long-range nature of the gravitational interaction.
 	Hence, there is no need to introduce a $3PV$ term as a correction term when Eq. 1 for the VME is used, we only need to account for the mass outside the particular boundary we are considering to evaluate the mass. An upper bound to the error when no knowleadge of the `real' extent of the cluster is also provided by (2).

\begin{figure}
\plotfiddle{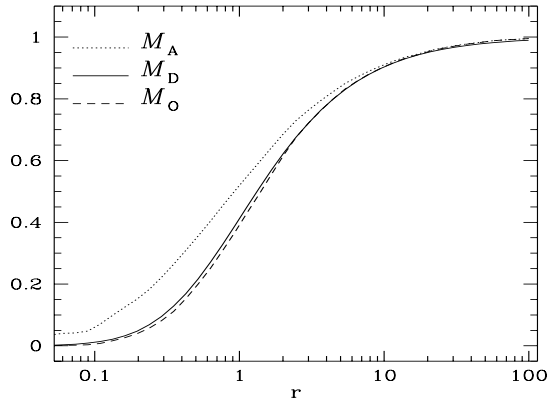}{5cm}{0}{40}{30}{-105}{-60}
\caption{Mass estimates using the aperture VME, $M_{\rm A}$, and that including the contribution to the potential energy from particles outside this aperture, $M_{\rm O}$. The true mass of the Dehnen model, $M_{\rm D}$, is also indicated. Here, the non-projected version of the VME is used, but the same holds for the projected quantity. }
\end{figure}

	When the physical boundary of a cluster is unknown the correction factor is model dependent.  Regarding real clusters, we are faced with a practical problem in astronomy. As we have seen, there is no problem for the VME to estimate the mass in an equilibrium system at different radii, provided that we actually know its total extent. But in astronomy, the physical boundary is a somewhat ambiguous matter, affected  by the same sensitivity of our instruments and even by the criteria used to `define' a cluster of galaxies. Nonetheless, one can use the difference in the mass estimated from the PME and from the VME to have an idea of the extent of the sampling done, as illustrated by Aceves \& Perea (1998) for the Coma cluster of galaxies.  This approach is more or less correct only when we can use other indicators, e.g.  X-rays, to infer that the cluster is near equilibrium. Further, the asymptotic approach to a particular value of the integrated mass is also indicative that the system is near complete sampling.

	In $N$-body simulations (e.g. Thomas \& Couchman 1992) the discrepancy seems to arise due to the use of formulae for the VME and the PME suitable only for test particles; i.e. not valid for mutually interacting particles. Furthermore, the possible presence of anisotropies in the outer parts of the numerical clusters can affect  the results especially those from the PME.

\section{An Anisotropic System}

In this section, as an illustrative example, we apply the virial theorem to an anisotropic system in equilibrium. The previous can give us an idea of the errors one might expect to commit under such conditions. For that purpose, we performed an $N$-body simulation of a cold collapse (e.g. van Albada 1982). The initial system had a ratio of twice the kinetic to potential energy of $2T/W = 0.1$, and the particles were distributed with a density law $\rho \propto 1/r\,$. We followed the evolution until the system reached a quasi-steady state, see Fig. 4.

\begin{figure}
\plotfiddle{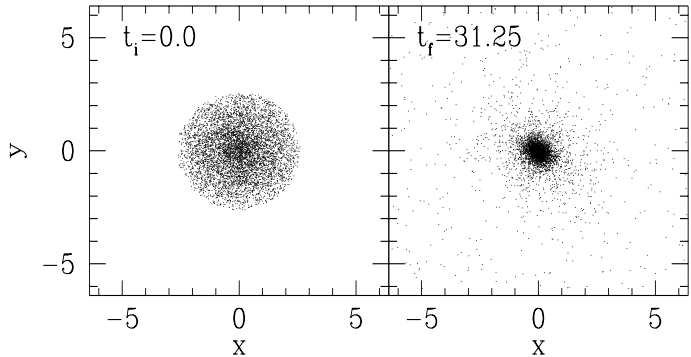}{4.5cm}{-90}{35}{43}{-130}{200}
\caption{Cold collapse of an initially $\propto 1/r$ distribution of matter, and $2T/W = 0.1\,$. ({\it Left}) The initial configuration of the system. ({\it Right}) System's appearance after having reached a quasi-steady state, $2T/W\approx 1$, at $t=31.25$ time units. The final system is ellipsoidal rather than spherical, and has a non-isotropic velocity distribution.}
\end{figure}

In Fig. 5 we show our results. On the left panel, the radial velocity dispersion and the projected velocity dispersion are shown for the final snapshot of the simulation; the anisotropy function $\beta(r)\equiv 1 - \sigma_{\theta}^2 / \sigma_r^2\,$ is shown in the inset. Note how the system is close to being isotropic, $\beta \approx 0$, at the inner regions while particles on radial orbits dominate the outer parts.

\begin{figure}
\plottwo{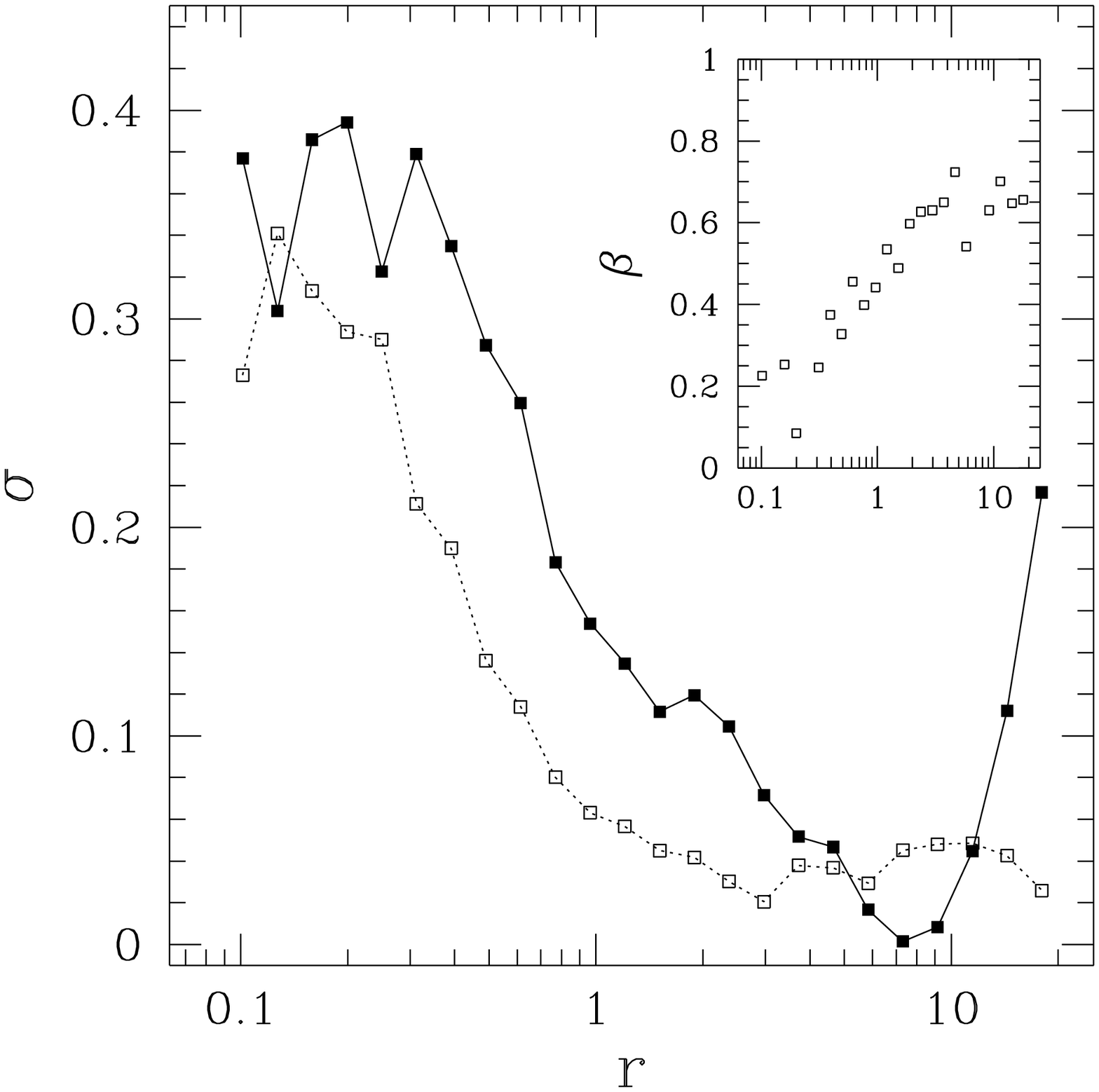}{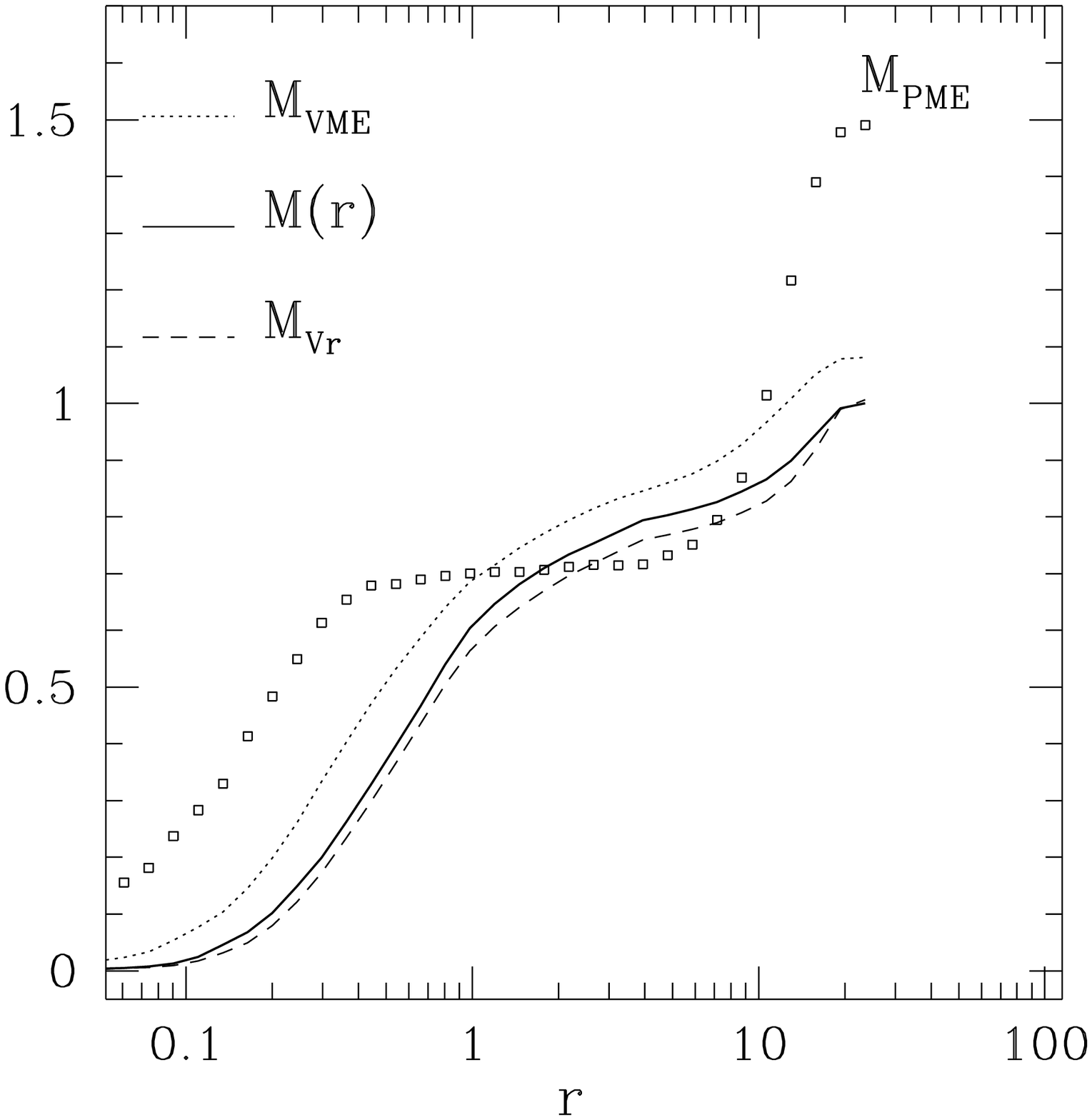}
\caption{({\it Left}) Radial velocity dispersion (filled squares), and projected velocity dispersion (open squares) in the final system of the cold collapse in Fig. 4. The inset depicts the anisotropy parameter $\beta(r)=1-\sigma_{\theta}^2/\sigma_r^2\,$. ({\it Right}) Application of the different mass estimators considered here. $M_{\rm Vr}$ corresponds to the non-projected form of the VME.}
\end{figure}

We applied the mass estimators described in $\S 2$ to the final configuration which is in a quasi-steady state. The PME and the VME were applied assuming isotropy in the system, i.e. formulae in Eq (1) were used, to reflect in a sense the way one proceeds in estimating the mass in clusters where this assumption is usually made; albeit observations indicating that clusters are somewhat triaxial (e.g. Plionis, et al. 1991). For the virial theorem estimates, we considered the contribution to the potential energy from the particles outside a given radius, and considered its projected version, $M_{\rm VME}$, and its non-projected one, $M_{\rm Vr}$. The true mass $M(r)$ was calculated by counting the particles inside spheres of different radii in the $N$-body system.

It is interesting to note that the non-projected virial theorem follows very close the true mass of the system, and the projected one overestimates it by $\approx 10$\% throughout most of the system's extent; see right panel of Fig. 5. This further supports the thesis that the VME is not very sensitive to anisotropies in equilibrium systems. We are not considering here the possible role of a mass-spectrum or rotation in clusters.

The virial theorem tends to be more sensitive to the equilibrium state of the system than to its anisotropy.  Since $M_{\rm Vr} \propto 2T/W$ it follows that for ``cold'' systems ($2T/W \la 1$)  the mass will be underestimated, while for ``hot'' systems ($2T/W \ga 1$) an overestimate will result. Indeed, for our initial configuration the mass estimate provided by the virial theorem was $M_{\rm Vr}(t=0)\approx 0.1\,$.

The PME is the most sensitive to anisotropies in the system, hence more propense to give unreliable mass estimates even for a quasi-equilibrium state.  Therefore, the VME provides a better estimate of the mass at different radii than the PME in situations where galaxies have non-isotropic velocity dispersions, and no information on the distribution of orbits is known.

\goodbreak

\section{Conclusions}

\begin{enumerate}

\item The projected mass estimator (PME) and the virial mass estimator (VME) yield accurate results for the total mass of a gravitational system, provided that the whole system is in steady-state, spherically symmetric, and has an isotropic velocity distribution.

\item When partial sampling is done, the error committed in the use of the PME and VME depends on the density profile, and their maxima occurs about the system's effective radius. For realistic profiles, the error can be up to $\approx 40$\% for the PME and $\approx 20$\% for the VME.

\item When the total extent of the system is known, the VME yields accurate mass estimates at different radii. This is achieved by accounting for the potential energy contributed from the particles outside a given radius.

\item A surface pressure correction term cannot be used in the mass 
calculated from the VME, when partial sampling of a system is done, to obtain a correct mass.  The correction term in this circumstance is dependent on the mass distribution of the cluster.

\item The VME tends to yield a mass profile closer to the true value in systems with anisotropic velocity distributions, provided this is in quasi-steady equilibrium.

\end{enumerate}

\acknowledgments

H.A. thanks the Spanish Ministry of Foreign Affairs for a MUTIS Scholarship. J.P. acknowledges support from the Spanish DGICYT Project PB96-0921. We also thank the Organizing Committee of the `Sesto 1998 Workshop on Observational Cosmology' for all their efforts to have a successfull and enjoyable meeting.


\end{document}